\documentclass[prl,twocolumn,superscriptaddress,showpacs]{revtex4}
\usepackage{graphicx,amsmath,amssymb,bm}

\def\be{\begin{equation}}
\def\ee{\end{equation}}
\def\ba{\begin{eqnarray}}
\def\ea{\end{eqnarray}}
\def\bas{\begin{eqnarray*}}
\def\eas{\end{eqnarray*}}

\begin{document}

\title{{\it Ab-initio} computation of neutron-rich oxygen isotopes}

\author{G.~Hagen} 
\affiliation{Physics Division, Oak Ridge National
Laboratory, Oak Ridge, TN 37831, USA} 
\author{T.~Papenbrock}
\affiliation{Department of Physics and Astronomy, University of
Tennessee, Knoxville, TN 37996, USA} 
\affiliation{Physics Division,
Oak Ridge National Laboratory, Oak Ridge, TN 37831, USA}
\author{D.J.~Dean} 
\affiliation{Physics Division, Oak Ridge National
Laboratory, Oak Ridge, TN 37831, USA} 
\author{M.~Hjorth-Jensen}
\affiliation{Department of Physics and Center of Mathematics for
Applications, University of Oslo, N-0316 Oslo, Norway}
\author{B. Velamur Asokan}
\affiliation{Computer Science and Mathematics Division, Oak Ridge National Laboratory,
Oak Ridge, TN 37831, USA}

\begin{abstract}
We compute the binding energy of neutron-rich oxygen isotopes and
employ the coupled-cluster method and chiral nucleon-nucleon
interactions at next-to-next-to-next-to-leading order with two
different cutoffs. We obtain rather well-converged results in model
spaces consisting of up to 21 oscillator shells. For interactions with
a momentum cutoff of 500~MeV, we find that $^{28}$O is stable with
respect to $^{24}$O, while calculations with a momentum cutoff of
600~MeV result in a slightly unbound $^{28}$O. The theoretical error
estimates due to the omission of the three-nucleon forces and the
truncation of excitations beyond three-particle-three-hole clusters
indicate that the stability of $^{28}$O cannot be ruled out from {\it
ab-initio} calculations, and that three-nucleon forces and continuum
effects play the dominant role in deciding this question.
\end{abstract}


\maketitle
{\it Introduction.} The neutron drip line marks the limits of
stability of neutron-rich isotopes. At present, this line is well
established only in the lightest elements, as the cross section for
the production of extremely neutron-rich nuclei decreases dramatically
as one moves away from nuclei in the valley of beta stability (for
a recent review see for example Ref.~\cite{Thoennessen04}). At present, $^{24}$O
is the ``last'' known stable neutron-rich oxygen ($Z=8$) isotope, and
$^{25}$O is known to decay under the emission of one
neutron \cite{Hoffman08}. The ``next'' neutron-rich oxygen isotope
$^{26}$O has not been observed
experimentally~\cite{Guillemaud90,Fauerbach96}, and systematics for
its production cross section suggest that it should have been seen if
it were a stable nucleus. Similar estimates suggest that the isotope
$^{28}$O is unstable~\cite{Tarasov97}. Thus, experiment puts the
neutron drip line at $^{24}$O. This is remarkable since $^{31}$F is
the most neutron-rich fluorine ($Z=9$) isotope~\cite{Sakurai99}. Thus,
the addition of a single proton apparently shifts the drip line by six
neutrons.

The theoretical determination of the neutron drip line is a
challenging task as well. Near the neutron drip line, small
uncertainties in the nuclear interaction are enhanced due to the
extreme isospin and the proximity of the continuum. Several
theoretical studies have addressed the structure of neutron-rich
oxygen and fluorine isotopes. The employed methods and theoretical
predictions differ considerably. The $sd$-shell model, based on the
Brown-Wildenthal USD interaction and the finite range droplet model,
predicts that $^{26}$O is stable~\cite{Brown05}. Within the $sd$-$pf$
shell model, the present experimental situation of an unstable
$^{26}$O can be reproduced after a modification of the
interaction~\cite{Utsuno02}. Within this model, $^{28}$O is unbound by
about 1~MeV. Within the same model space, but a different interaction,
Poves and Retamosa~\cite{Poves94} obtained a stable $^{31}$F and a
stable $^{28}$O. Shell-model descriptions of neutron-rich oxygen
isotopes, including the coupling to the continuum, were given in
Refs.~\cite{Luo02,Michel,Volya06}. Within the latter
approach~\cite{Volya06}, two slightly different phenomenological $sd$-shell
interactions are employed for oxygen isotopes close to and far away
from the valley of beta stability, respectively. This leads to the
result that $^{26}$O is unstable with respect to $^{24}$O, while
$^{28}$O is unstable with respect to the emission of two and four
neutrons. Clearly, the present theoretical situation does not have the
desired predictive power, and calculations suffer from uncertainties
in the knowledge of the interaction and from the difficulty to
quantify how these uncertainties propagate in the quantum many-body
problem. This is an opportunity for {\it ab-initio} calculations to 
address these challenges.

{\it Ab-initio} calculations have been very successful in light
nuclei~\cite{Piep01,Nav00,Kam01,Fujii04,Nav07}, and have recently also
been extended to unbound~\cite{Hag06,Noll06,Quag08} and
medium-mass isotopes~\cite{Hag08}.  In this paper we present {\it
ab-initio} calculations for the neutron-rich oxygen isotopes
$^{22,24,28}$O, and employ nucleon-nucleon interactions from chiral
effective field theory
(EFT)~\cite{Weinberg,Kolck94,Epel00,N3LO,N3LOEGM}. These interactions
are rooted in quantum-chromo-dynamics and include pion exchange and
zero-range forces. The power counting, i.e., the systematic expansion
of the interaction in terms of ratios of the probed momentum scale $Q$
over the cutoff $\Lambda_\chi$ is an important asset. In finite
nuclei, $Q$ is about 200~MeV~\cite{Hag08}, while the cutoffs we employ
are $\Lambda_\chi=500$~MeV and $\Lambda_\chi=600$~MeV,
respectively. The variation of our results with the cutoff allows us
to quantify uncertainties that are due to the omission of
(short-ranged) three-nucleon forces. We employ the
coupled-cluster method~\cite{Coe58,Coe60,Ciz66,Ciz69,Kuem78,Bar07} for
the solution of the quantum many-body problem. This method scales
gently with the system size,  
and can accurately compute the binding energies of nuclei with closed
subshells. In particular, the possibility to employ large model spaces
avoids the need for a secondary renormalization of the chiral
interactions from EFT for nuclei such as oxygen and calcium
isotopes~\cite{Hag08}. 

This paper is organized as follows. We briefly
introduce the interactions and methods we employ and then present the
results of our calculations. 

{\it Interaction, model space and coupled-cluster method}: We employ
the chiral nucleon-nucleon interaction by Entem and
Machleidt~\cite{N3LO} at next-to-next-to-next-to-leading order
(N$^3$LO). This includes terms up to order $(Q/\Lambda_\chi)^4$ in the
power counting of the nucleon-nucleon interaction. The interaction has
a high-momentum cutoff of $\Lambda_\chi=500$~MeV, and a version with
cutoff $\Lambda_\chi=600$~MeV is also available \cite{N3LO600}. The low-energy
constants of the chiral potentials were determined by fits to the
two-nucleon system. We neglect three-nucleon forces that already
appear at next-to-next-to-leading order and thereby introduce
uncertainties of the order $(Q/\Lambda_\chi)^3$. As physics must be
independent of the cutoff (or renormalization scale), any
cutoff-dependence in our results quantifies the uncertainty due to
omitted contributions of short-ranged three-nucleon forces and forces of higher
rank. The intrinsic Hamiltonian is
\be
\label{ham}
\nonumber
\hat{H} = \hat{T}-\hat{T}_{\rm cm} +\hat{V}(\Lambda_\chi) \ .
\ee
Here $\hat{T}$, $\hat{T}_{\rm cm}$, and $\hat{V}(\Lambda_\chi)$ denote
the kinetic energy of the $A$-body system, the kinetic energy of the
center-of-mass coordinate, and the chiral nucleon-nucleon interaction
with momentum cutoff $\Lambda_\chi$, respectively. The intrinsic
Hamiltonian~(\ref{ham}) is translationally invariant and does not
depend on the center-of-mass coordinate. We express the Hamiltonian in
a single-particle basis of the spherical harmonic oscillator. Our
model-space parameters are the oscillator spacing $\hbar\omega$ of our
single-particle basis, and the maximal excitation energy
$(N+3/2)\hbar\omega $ of a single-particle state, that is the number of
complete oscillator shells is $N+1$. As a first step towards the
solution of the many-body problem, we solve the spherical Hartree-Fock
equations and transform the Hamiltonian to this basis. In the second
step, we employ the coupled-cluster method. In drip-line nuclei, the
outermost nucleons move in orbitals close to the scattering
threshold, making the nuclear wave function exhibit halo-like structures
and sometimes even ground states embedded in the continuum. 
The presence of the scattering continuum in such exotic nuclei makes the
use of the oscillator basis not ideal. The unrealistic Gaussian falloff of the 
oscillator wave functions makes convergence slow for nuclei with 
dilute matter distributions. 
However, the Gamow-Hartree-Fock (see, e.g., Ref.~\cite{Hag06})
yields occupied single-particle states with 
nonphysical positive imaginary parts. This difficulty
is due to the relatively ``hard'' interaction we employ.
To avoid this problem, we choose to stay within the oscillator basis, but employ
very large model spaces for an improved description of the tails of the
radial wave function.

The nuclear many-body problem is solved with the coupled-cluster
method~\cite{Coe58,Coe60,Ciz66,Ciz69,Kuem78,Bar07,Hei99,Dean04,
Kow04,Wlo05}. This approach is based on the similarity transformation 
of the normal-ordered intrinsic Hamiltonian $\hat{H}_{N}$, 
\be
\label{hbar}
\overline{H}=e^{-\hat{T}} \hat{H}_N e^{\hat{T}} \ .
\ee
Here, the Hamiltonian is
normal-ordered with respect to a product state $|\psi\rangle$ which
serves as a reference. The particle-hole cluster operator
\begin{equation}
\label{T}
\hat{T} = \hat{T}_1 + \hat{T}_2 + \hat{T}_3+\ldots + \hat{T}_A
\end{equation}
is defined with respect to this reference state. It is a sum of the $k$-particle 
$k$-hole ($k$p-$k$h) cluster operators 
\begin{equation}
\hat{T}_k =
\frac{1}{(k!)^2} \sum_{i_1,\ldots,i_k; a_1,\ldots,a_k} t_{i_1\ldots
i_k}^{a_1\ldots a_k}
\hat{a}^\dagger_{a_1}\ldots\hat{a}^\dagger_{a_k}
\hat{a}_{i_k}\ldots\hat{a}_{i_1} \ .
\end{equation}

We use the convention that $i, j, k,\ldots$ label occupied
single-particle orbitals while $a,b,c,\ldots$ label unoccupied
orbitals. We truncate the cluster operator beyond the $\hat{T}_2$ level and
employ $\Lambda$CCSD(T)~\cite{Kuch,Taube} as an approximation for 
the $\hat{T}_3$ clusters.
The unknown cluster amplitudes $t_i^a$ and $t_{ij}^{ab}$
in Eq.~(\ref{T}) are determined from the solution of the coupled-cluster 
equations
\be
\label{ccsd}
0 = \langle \psi_i^a | \overline{H} | \psi\rangle \ , \qquad
0 = \langle \psi_{ij}^{ab} | \overline{H} | \psi\rangle \ .
\ee
Here $|\psi_i^a\rangle = \hat{a}_{a}^\dagger\hat{a}_{i}|\phi\rangle$
and $|\psi_{ij}^{ab}\rangle =
\hat{a}_{a}^\dagger\hat{a}^\dagger_b\hat{a}_j\hat{a}_{i}|\psi\rangle$
are 1p-1h and 2p-2h excitation of the reference state, respectively.

The nonlinear coupled-cluster equations~(\ref{ccsd}) are solved
iteratively, and the correlation energy of the ground state is
computed from
\be 
\Delta E_{\rm CCSD} = \langle \psi | \overline{H} | \psi\rangle \ .  
\ee 

We employ a spherical formulation of coupled-cluster theory where the
cluster operator $\hat{T}$ is a scalar under
rotations~\cite{Hag08}. This formulation reduces considerably the
number of unknowns and permits us to explore model spaces exceeding 20
major oscillator shells.  

Let us briefly summarize the essential
properties of the coupled-cluster method. First, the 
method fulfills Goldstone's linked cluster theorem and therefor
yields size-extensive results, i.e., the error due to the truncation is
linear in the mass number $A$. Size extensivity is an important issue when approximate
solutions to all but the lightest nuclei are
sought~\cite{Bar07,comment}. Second, the computational effort scales
gently (i.e., polynomial) with the system size. The method has met
benchmarks in light nuclei~\cite{Mi00,CCbench}. We neglect
three-nucleon forces since their application within the
coupled-cluster method is presently limited to smaller model
spaces~\cite{Hag07}.

For a more precise computation of the correlation energy, we consider 
corrections due to triples excitations $\hat{T}_3$ within
the $\Lambda$CCSD(T) approximation. For this purpose, we solve the left
eigenvalue problem 
\be
\label{left}
\langle\psi| \hat{\Lambda} \overline{H} = E\langle\psi| \hat{\Lambda}
\ee
of the similarity-transformed Hamiltonian $\overline{H}$. Here, $\hat{\Lambda}$ denotes
the de-excitation cluster operator
\be
\hat{\Lambda} = 1 + \hat{\Lambda}_1 +\hat{\Lambda}_2 \ ,
\ee
and 
\be
\hat{\Lambda}_1 =
\sum_{i,a} \lambda^{i}_{a}
\hat{a}_{a}
\hat{a}^\dagger_{i} \ , \quad
\hat{\Lambda}_2 =
\frac{1}{4} \sum_{i,j,a,b} \lambda^{ij}_{ab}
\hat{a}_{b}\hat{a}_{a}
\hat{a}^\dagger_{i}\hat{a}^\dagger_{j} \ .
\ee
The unknowns $\lambda^i_a$ and $\lambda^{ij}_{ab}$ result from the ground-state
solution of the left eigenvalue problem (\ref{left}). They are
utilized together with the cluster amplitudes $t_i^a$ and
$t_{ij}^{ab}$ to compute the energy correction due to triples clusters
as
\ba
\Delta E_3&=& {1\over (3!)^2} \sum_{i j k a b c} \langle\psi|
\hat{\Lambda}(\hat{F}_{hp} +\hat{V})_N|\psi^{abc}_{ijk}\rangle\nonumber\\
&\times& {1\over
\varepsilon_{ijk}^{abc}}\langle\psi_{ijk}^{abc}|(\hat{V}_N\hat{T}_2)_C|\psi\rangle
\ .  
\ea

Here, $\hat{F}_{hp}$ denotes the part of the normal-ordered one-body
Hamiltonian that annihilates particles and creates holes, while 
$\varepsilon_{ijk}^{abc} \equiv f_{ii}+f_{jj}+f_{kk}-f_{aa}-f_{bb}-f_{cc}$
is expressed in terms of the diagonal matrix elements of the
normal-ordered one-body Hamiltonian $\hat{F}$.  The subscript $C$
denotes the connected part of the operator, and
$|\psi_{ijk}^{abc}\rangle$ is a 3p-3h excitation of the reference
state.
  
\begin{figure}[h]
\includegraphics[width=0.35\textwidth,clip=]{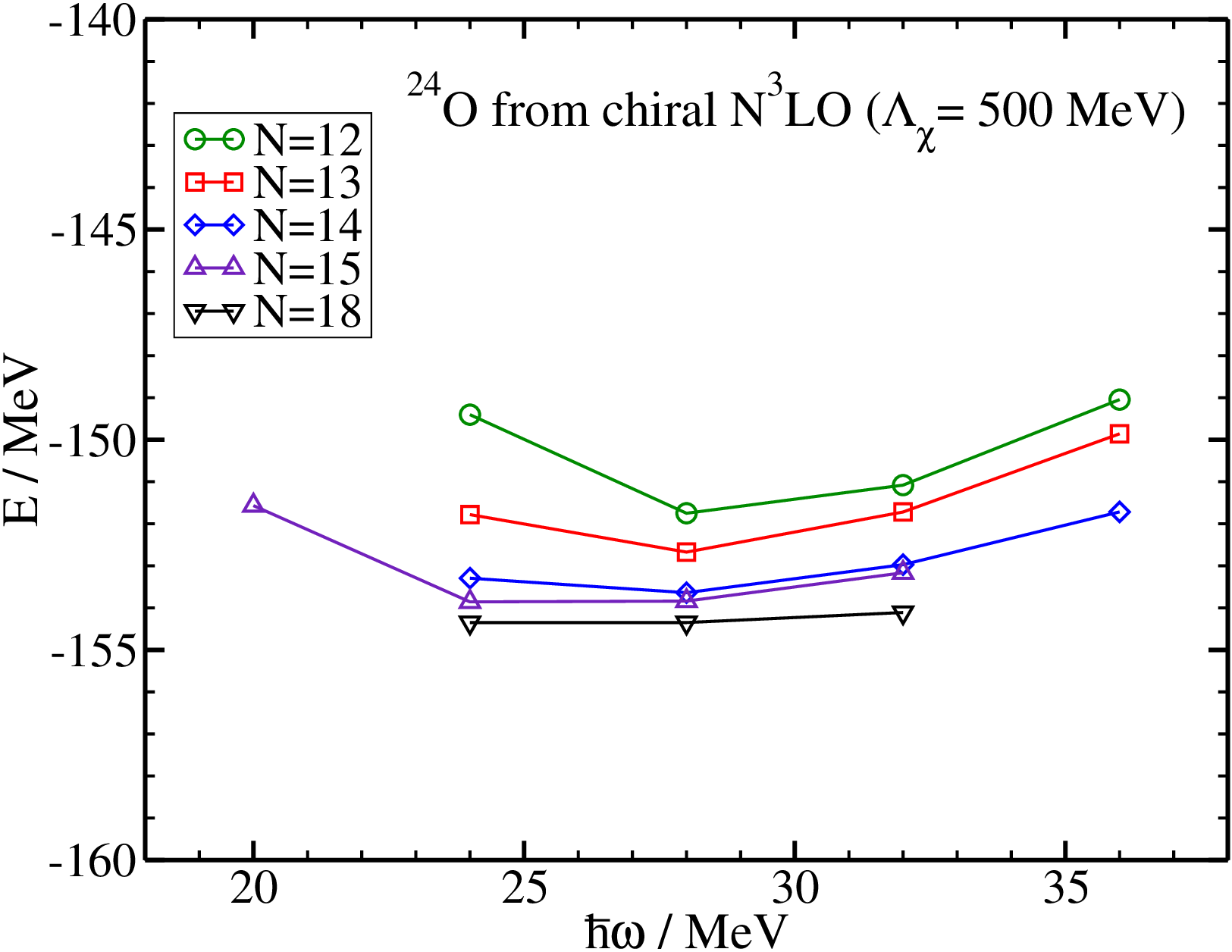}
\caption{(Color online) Binding energy (within $\Lambda$CCSD(T)) for
$^{24}$O from a chiral NN potential at order N$^3$LO with
high-momentum cutoffs $\Lambda_\chi=500$~MeV as a function of the
oscillator spacing $\hbar\omega$ and the size of the model space.}
\label{fig1}
\end{figure}

\begin{figure}[h]
\includegraphics[width=0.35\textwidth,clip=]{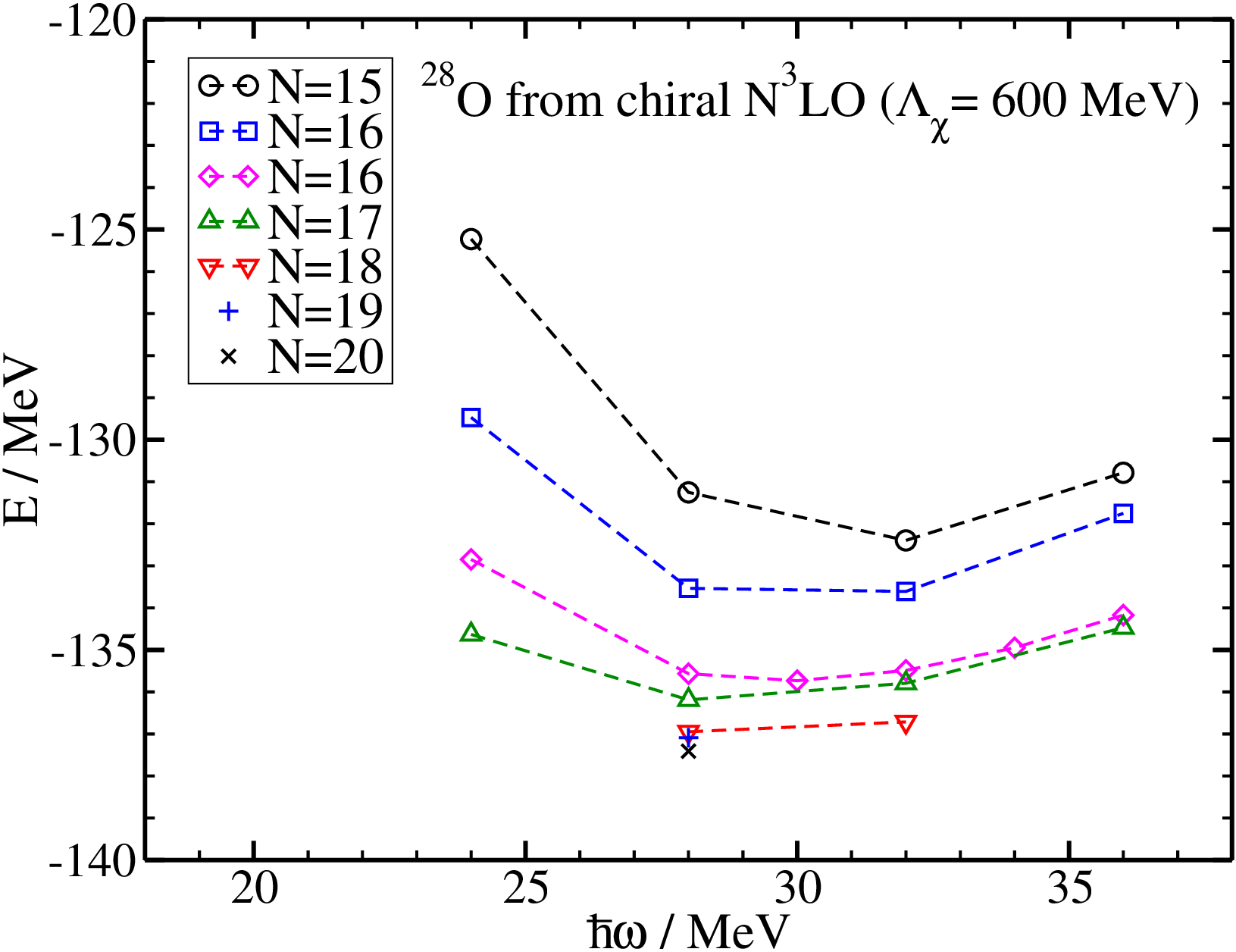}
\caption{(Color online) Same as Fig.~\ref{fig1} except for $^{28}$O and a
chiral cutoff $\Lambda_\chi=600$~MeV.}
\label{fig2}
\end{figure}

{\it Results:} We consider the nuclei $^{16,22,24,28}$O and compute
their ground-state energies within the $\Lambda$CCSD(T) approximation
for chiral interactions with cutoffs of $\Lambda_\chi=500$~MeV and
$\Lambda_\chi=600$~MeV, respectively. Figures~\ref{fig1} and
\ref{fig2} show the results as a function of the oscillator spacing
$\hbar\omega$ of the single-particle basis, and parametrized by the
number of major oscillator shells $N+1$ for $^{24}$O, and $^{28}$O
with two chiral cutoffs $\Lambda_\chi$, respectively. Note that the
results are reasonably well converged with respect to the size of the model
space. Note also that the ``harder'' interaction with cutoff
$\Lambda_\chi=600$~MeV requires a larger model space to reach an
acceptable convergence. The results for $^{16}$O and $^{22}$O are of
similar quality. Let us also comment on the separation of the
center-of-mass coordinate and the intrinsic coordinates.  Very
recently, Hagen, Papenbrock, and Dean demonstrated that the
coupled-cluster wave function is approximately a product of a
translationally invariant wave function and a Gaussian for the
center-of-mass coordinate~\cite{HPD}. Thus, we do not worry
about spurious contributions to the coupled-cluster wave function.

Let us estimate the precision of our results. There are three sources
of systematic errors, namely the truncation level of the
coupled-cluster method, the finite size of the model space, and the
error due to omitted contributions in the interaction.  First, within
the $\Lambda$CCSD(T) approximation, three-particle-three-hole clusters are treated
approximately, and all excitation clusters of higher rank are
neglected. Table~\ref{tab1} shows the different contributions to the
binding energy of neutron-rich oxygen isotopes. Comparison of the CCSD
correlation energy and the energy due to triples corrections shows that
the latter account for 10\% (13\%) of the former at a cutoff
$\Lambda_\chi=500$~MeV ($\Lambda_\chi=600$~MeV). These ratios are
found in similar coupled-cluster calculations of atoms and
molecules, and experience in quantum chemistry (see for example
Ref.~\cite{Bar07}) suggests that the truncation of the cluster
amplitudes beyond the triples corrections introduces an error of a few
percent corresponding to an uncertainty of approximately 5~MeV.  Second, we cannot
treat an infinite model space, and (as shown in Figs.~\ref{fig1} and
\ref{fig2}), the convergence with respect to an increased size of
the model space is at the level of a couple of MeV. Thus, the
convergence with respect to the size of the model space introduces an
error that does not exceed error estimates due to the truncation of
the cluster amplitudes. Third, by far the largest uncertainty is due
to omissions in the nuclear interaction, as can be seen from a
comparison of the results obtained with two different cutoffs. This
uncertainty is of the order of 10--20~MeV, and increases with
increasing mass number.  Note that the deviation from the experimental
results is consistent with our error estimates. Overall, we are
missing binding energy compared to experiment. Thus, the
net effect of the three-nucleon force is expected to be attractive for
both cutoffs. Table~\ref{tab2} gives the root-mean-square point 
matter radii using the chiral interaction with high-momentum cutoff
$\Lambda_\chi = 500$MeV. 

Radii were calculated 
using the Helmann-Feynman theorem within the $\Lambda$CCSD(T)
approximation in 19 major oscillator shells. We compare with the 
effective point matter radii extracted from interaction cross sections
using the Glauber model in the optical limit approximation
\cite{Ozawa01}. Our calculated matter radii are smaller than those
extracted from experiment. In our calculated radii we estimate an
uncertainty at the order of  $\sim 0.1$ fm from the model-space
dependence. 

Let us also check whether our error estimates are consistent with the
power-counting estimates from chiral EFT. Nogga confirmed that these
estimates hold in light nuclei~\cite{Nogga06}. The omitted
three-nucleon forces are of the order $\langle \hat{V}\rangle
(Q/\Lambda_\chi)^3$ where $Q$ is the typical momentum scale and
$\langle \hat{V}\rangle$ is the expectation value of the two-body
interaction. For nuclei in this mass region and a cutoff
$\Lambda_\chi=500$~MeV, we have $Q\approx 200$~MeV and $\langle
\hat{V}\rangle\approx 33\pm 3$~MeV per nucleon (taken from the
expectation values of the kinetic and potential energies in
Ref.~\cite{Hag08}, respectively). This puts power-counting estimates
from chiral EFT at about 2~MeV per nucleon, and our results are well
within this estimate. While the absolute uncertainty on the binding
energy is thus considerable, the differences in the binding energies
of the considered isotopes (at fixed chiral cutoff $\Lambda_\chi$) is
much closer to the experimental result.

\begin{table}[h]
\begin{tabular}{|l||r|r|r|r|}\hline
  Energies          & $^{16}$O   &$^{22}$O & $^{24}$O& $^{28}$O\\\hline\hline
($\Lambda_\chi=500$~MeV)&       &         &         &         \\ 
$E_0$              &   25.946 &  46.52   &  50.74   &  63.85  \\
$\Delta E_{\rm CCSD}$ & -133.53  & -171.31 & -185.17 & -200.63\\
       $\Delta E_3$&   -13.31  &  -19.61 &  -19.91 &  -20.23 \\
       $E$         &     -120.89  &  -144.40  &  -154.34 &  -157.01 \\
($\Lambda_\chi=600$~MeV)&       &         &         &         \\
$E_0$              &      22.08 &   46.33 &   52.94 &   68.57\\
$\Delta E_{\rm CCSD}$ & -119.04 & -156.51 & -168.49 & -182.42\\
$\Delta E_3$       &     -14.95 &  -20.71 &  -22.49 &  -22.86 \\
       $E$         &    -111.91 & -130.89 & -138.04 & -136.71 \\\hline
Experiment         &    -127.62 & -162.03 & -168.38 &        \\ \hline
\end{tabular}
\caption{Contributions to the binding energy $E$ (in MeV) in neutron-rich
oxygen isotopes from chiral interactions with high-momentum cutoff
$\Lambda_\chi$. The contributions $E_0$, $\Delta E_{\rm CCSD}$, and $\Delta
E_3$ denote the Hartree-Fock energy, the correlation energy within the
CCSD approximation, and the energy due to the employed triples
correction, respectively.  For 
$^{16}$O and $\Lambda_\chi = 500$MeV the results were
taken in 19 major oscillator shells and at the energy minimum
$\hbar\omega = 40$MeV. For all other cases the results were obtained in the
largest model spaces at fixed $\hbar\omega=28$~MeV.}
\label{tab1}  
\end{table}

\begin{table}[h]
\begin{tabular}{|l||l|l|l|l|}\hline
            & $^{16}$O   &$^{22}$O & $^{24}$O& $^{28}$O\\\hline\hline
            $\langle r^2\rangle^{1/2} $  & 2.296  & 2.405  & 2.658  &
            2.825 \\
            \hline
            Expt. & 2.54(2) & 2.88(6) & 3.19(13) &    \\\hline 
\end{tabular}
\caption{Root-mean-square point matter radii (in fm) for neutron-rich
oxygen isotopes from the chiral interaction with high-momentum cutoff
$\Lambda_\chi = 500$MeV. Oscillator frequencies as in Table~\ref{tab1}.
Experimental data from Ref.~\cite{Ozawa01}.}
\label{tab2}  
\end{table}

\begin{figure}[h]
\includegraphics[width=0.35\textwidth,clip=]{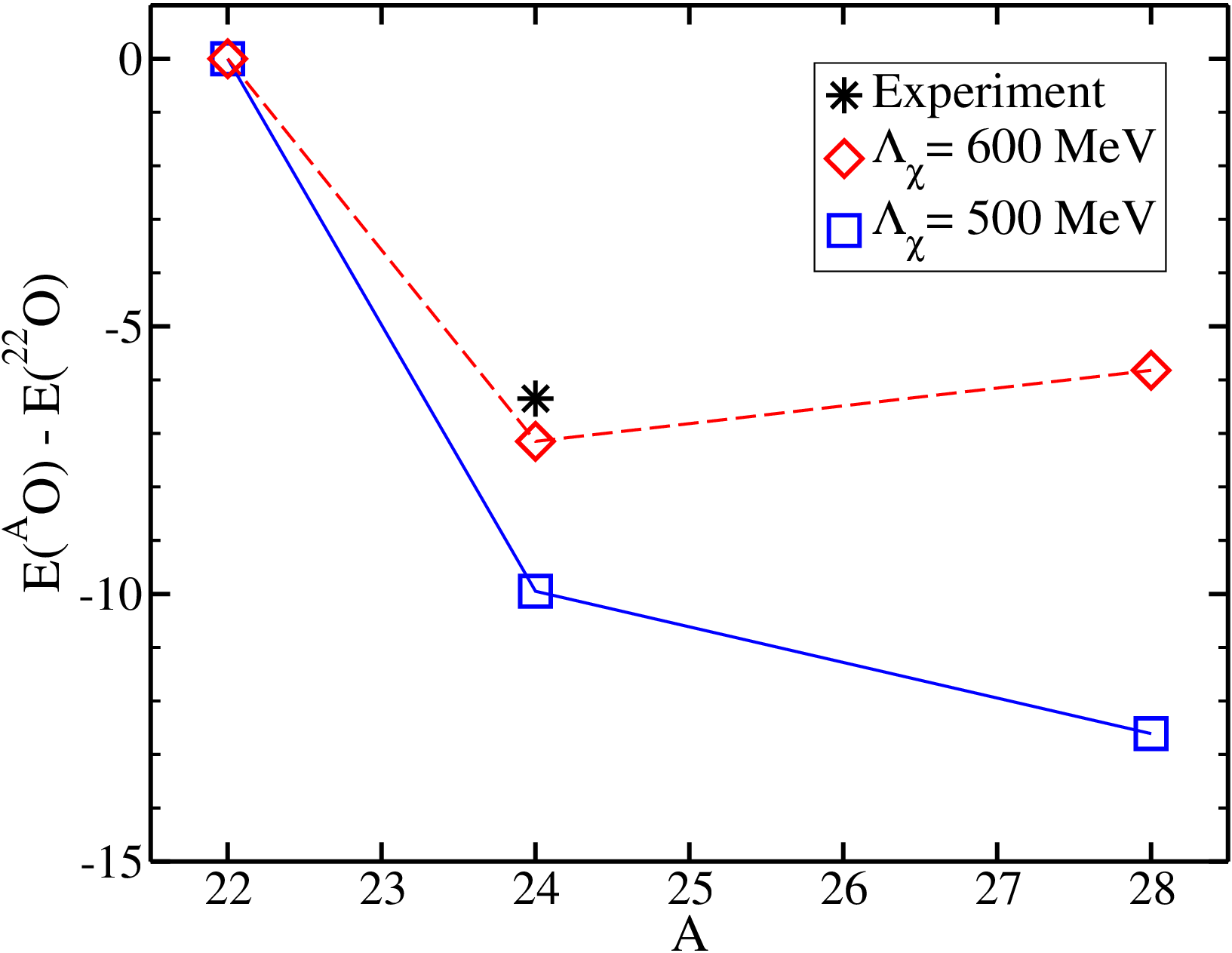}
\caption{(Color online) Ground-state energies of neutron-rich oxygen
isotopes $^A$O relative to $^{22}$O for chiral
interactions with two different cutoffs $\Lambda_\chi$.}
\label{fig3}
\end{figure}

Let us focus on the binding energy of $^{28}$O with respect to
$^{24}$O. While it would certainly be interesting to include $^{26}$O
in this comparison, we cannot address this nucleus within the
spherical coupled-cluster method due to its open-shell
character. Recall, however, the experimental
evidence~\cite{Guillemaud90,Fauerbach96} against the stability of
$^{26}$O. This makes the comparison of $^{28}$O and the last known
stable isotope $^{24}$O particularly interesting. Figure~\ref{fig3} shows
that the ground-state energies relative to $^{22}$O change little as
one goes from $^{24}$O to $^{28}$O. This is a remarkable result of our
{\it ab-initio} calculations. In shell-model calculations with an
$^{16}$O core, the ground-state energies typically increase  strongly (in absolute value)
as more neutrons are added, and an adjustment of the interaction is
necessary~\cite{Volya06}. We study this phenomenon further by
employing a similarity renormalization group (SRG)
transformation~\cite{SRG} of the Hamiltonian with cutoff
$\Lambda_\chi=500$~MeV. As we lower the smooth SRG momentum cutoff
from 4.1~fm$^{-1}$ to 3.5~fm$^{-1}$, we find that the ground-state
energy of $^{28}$O decreases further relative to $^{24}$O. Thus, a
softening of the nucleon-nucleon interaction has to be compensated by
a three-nucleon force that yields less attraction (or even repulsion) in
$^{28}$O than in $^{24}$O. There is no cutoff in this range that
simultaneously would reproduce the experimental binding of $^{22}$O
and $^{24}$O.

At a cutoff $\Lambda_\chi=500$~MeV we find that $^{28}$O is bound by
about 2.7~MeV with respect to $^{24}$O.  However, the situation is
reversed at the higher cutoff $\Lambda_\chi=600$~MeV, and the
difference is about -1.3~MeV. Given the uncertainties of our
calculation as discussed in the preceding paragraph, it is presently
not possible to reach a conclusion regarding the existence of
$^{28}$O.  However, the discussion presented in the previous paragraph
also makes clear that -- within interactions from chiral EFT -- the
stability of $^{28}$O depends mainly on the contributions of the
three-nucleon force, and that even small contributions can tip the
balance in either direction. This is the main result of this
paper. Our {\it ab-initio} calculations also suggest that the recent results
from phenomenological shell-model approaches regarding the unbound
character of $^{28}$O might be viewed with caution. The combination
of three-nucleon forces, the proximity of the continuum and the
isospin dependence are a challenge for reliable theoretical predictions.

In summary, we performed {\it ab-initio} calculations for neutron-rich
oxygen isotopes employing chiral nucleon-nucleon interactions at order
N$^3$LO. We probed the effects of missing physics (such as
three-nucleon forces) by studying the cutoff dependence of our
results, and estimated the uncertainties due to the finite size of the
model space and the truncation of the cluster operator. Our results
show that the absolute binding energies have considerable
uncertainties. However, the differences in binding energies are much
closer to experiment. We find a small difference in
the binding energies of $^{24}$O and $^{28}$O. Thus, our results
cannot rule out a stable $^{28}$O with respect to $^{24}$O. The
cutoff-dependence of the results shows that three-nucleon forces are
the dominant contributions that tip the balance.

We thank P. Fallon, R. Furnstahl, K. Jones, M. P{\l}oszajczak,
A. Schiller and A. Taube for useful discussions.  This work was
supported by the U.S. Department of Energy under Contract Nos. \
DE-AC05-00OR22725 with UT-Battelle, LLC (Oak Ridge National
Laboratory), and DE-FC02-07ER41457 (SciDAC UNEDF), and under Grant
No.\ DE-FG02-96ER40963 (University of Tennessee).  This research used
computational resources of the National Institute for Computational
Sciences (UT/ORNL) and the National Center for Computational Sciences
(ORNL).

\end{document}